# From Raw Gaze to Meaningful Features: Assessment of Visual Behavior in Driving Simulator

Smilja Stokanović[1*], Jaka Sodnik[2], Nadica Miljković[1,2]

1: University of Belgrade - School of Electrical Engineering, Bulevar kralja Aleksandra 73, 11000 Belgrade, Serbia

2: Faculty of Electrical Engineering, University of Ljubljana, Tržaška cesta 25, 1000 Ljubljana, Slovenia

*Corresponding Author**

*e-mails:* *smiljastokanovic@gmail.com*, *jaka.sodnik@fe.uni-lj.si*, *nadica.miljkovic@etf.bg.ac.rs*

## Abstract

A novel quantitative analytical approach to investigate eye movement behaviour in a driving simulator during three conditions: Baseline, Ride (simulated drive under normal visibility), and Fog (simulated drive under reduced visibility) is presented. We illustrate the proposed approach in a case study: eye tracking data from 24 participants are analyzed using 31 parameters, organized into three groups: (1) saccade characteristics, (2) Bivariate Contour Ellipse Area (BCEA) related parameters, and (3) blinking features. Across all feature groups, numerous statistically significant differences emerge between Baseline and the simulated drive conditions. Between Ride and Fog, saccade features show minimal changes (one out of 13), whereas BCEA (9 of 13) and blink features (four of 5) exhibit pronounced differences, highlighting the strong impact of reduced visibility on gaze stability and blinking behaviour. In addition to conventional gaze features, such as saccade duration, a new metric, inspired by the Guzik's Index (GI), is introduced through feature engineering to quantify fixation asymmetry along the major axis of the BCEA. The GI results suggest that gaze dispersion remains consistent across visibility conditions, but it is significantly higher during Baseline than during Fog and Ride conditions ($p < 0.01$). Overall, the presented findings demonstrate that a combination of the BCEA-based features with saccade and blink parameters provides a comprehensive understanding of visual attention and gaze stability in a driving simulator, while GI offers additional insights into fixation asymmetry under varying visibility conditions.

# Introduction

Driving represents a complex, multitasking activity that demands both mental acuity and physical coordination (Regan *et al.,* 2008; Guidetti *et al.,* 2019). Visual attention plays a critical role in safe driving, with eye movements serving as a key indicator of cognitive and perceptual processing (Guidetti *et al*., 2019; Borowsky *et al*., 2015; Liang and Lee, 2010; Zhang *et al.,* 2014). Eye movements are crucial for decision-making, which is pivotal for safe driving (Orquin and Loose, 2013; Mackenzie and Harris, 2015; Crundall and Underwood, 2011). Particularly, rapid eye movements or saccades, which can reach the speed of up to 500 degrees per second and typically last less than 100 ms (Leigh and Zee, 2015), allow drivers to acquire precise visual information quickly (Guidetti *et al.,* 2019). Saccades are essential for scanning the environment, detecting hazards, and making decisions (Yang *et al.,* 2021; Barrett and Elliott, 2011; Zirnsak and Moore, 2015; Ting and Gluth, 2024; Spering, 2022) - all being crucial for assessing situation awareness and driver's behavior (Endsley, 2020). The rapid nature of saccadic movements presents a particular challenge for gaze-based interface design, requiring analytical approaches beyond parameters derived from manual input devices (Porta, 2002). Luckily, saccades can be easily detected and analyzed using eye tracking technology, providing detailed insight into visual attention and gaze behavior during driving in a relatively unobtrusive manner (Stuart *et al.,* 2019; Madriaga *et al.,* 2023).

## Motivation for Introducing Dedicated Eye Tracking Features

Although fog presents a major crash risk factor (Wu *et al.,* 2018; Abdel-Aty *et al.,* 2011), most eye tracking research focused on investigation under clear conditions (Benedetto *et al.,* 2011*;* Kujala *et al.,* 2016*;* Deng *et al.,* 2020; Vetturi *et al.,* 2020), leaving a critical gap in understanding how oculomotor behavior adapts to reduced visibility. Current driver monitoring systems rely on single metric approaches, measuring either saccade patterns, fixation behavior, or blink rate, which have proved useful for detecting fatigue or distraction under normal visibility conditions (Recarte *et al., 2008;* Benedetto *et al.,* 2011). Drivers in fog may suppress blinking to maintain visual information and alter both saccade patterns and gaze dispersion simultaneously - changes that unimodal systems failed to capture comprehensively (Nishizono *et al.,* 2023; Lian *et al.,* 2024). This gap is particularly problematic for Human-Computer Interface (HCI) design, as Advanced Driver-Assistance System (ADAS) requires robust indicators capable of distinguishing normal driving and visual strain caused by poor visibility (Hills, 1980).

Eye tracking offers key practical advantages over other physiological sensors: it is relatively non-intrusive, requires no skin contact, it is compatible with corrective lenses, and can be quickly deployed and synchronized (Kapitaniak *et* al., 2015; Shuetz and Fiehler, 2022; Nyström *et al.,* 2013; Bergasa *et al.,* 2006). Our goal is to determine whether eye tracking alone, as a single sensor, can efficiently detect different visual exploration patterns across varying environmental demands. The key question is not simply whether to use eye tracking, but how to extract meaningful patterns from gaze data that reliably reflect driver adaptation to environmental challenges.

To address this, we propose a multiparameter, comprehensive, and effective approach integrating three groups of features obtained from eye movement measurements. Saccade detection provides insights into eye movement dynamics and visual exploration strategies. The Bivariate Contour Ellipse Area (BCEA) related parameters serve as a spatial measure of gaze dispersion, reflecting the stability of visual attention. Eye blinks, as indicators, provide the third dimension of driver monitoring, as blink patterns can indicate changes in processing demands, search efficiency, and potential fatigue or overload (Faure *et al.,* 2016). Moreover, a newly introduced feature, Guzik's Index (GI), may capture directional asymmetries or instability in fixation behavior that conventional summary statistics may overlook, thereby increasing the sensitivity and diagnostic power of eye movement analysis. A novel aspect of this work is the systematic integration of multiparameter assessment under reduced visibility, with possible implications for driving evaluation and adaptive HCI.

## Research Objectives

The central question guiding this work is: How do eye movements differ during the resting phase and simulated driving (with changed visibility), and can we detect changes in three distinct measurement conditions by using eye tracker data? The key research questions are:

1. Does incorporating BCEA-based features (a novel approach in driving research) alongside saccade and blink measures enhance understanding about visual attention and gaze stability during Baseline and simulated driving tasks under normal and altered visibility?
2. Can eye tracker data reveal distinct changes across three measurement conditions?
3. How does the presence of fog affect visual attention and gaze stability in comparison to the Ride and Baseline conditions?
4. To what extent does the newly introduced parameter - GI, provide additional insights into eye movement dynamics and visual attention during simulated driving?

To address research questions, we present an analytical strategy for analyzing eye tracker data that integrates BCEA related parameters with widely used oculomotor measures, such as saccades and blinks, to better understand visual attention and gaze stability during simulated driving conditions. In total, three categories of parameters are analyzed, comprising an extensive list of 31 parameters overall: saccade-related parameters (SP) capturing rapid eye movements, BCEA-based features assessing gaze variability (BCEA), and blink-related parameters (BP) reflecting eye closure patterns. Overall, the study aims to determine whether multi-modal eye tracking parameters can serve as an effective metric that enhances the interpretation of visual behavior. Understanding these oculomotor adaptations is critical for driving safety evaluation and HCI development.

# Materials and Methods

In this study, eye tracker data are collected, preprocessed, and relevant features are extracted. Based on the extracted parameters, statistical analysis is performed to examine differences and relationships between the three conditions - Baseline, Ride (simulated driving under normal visibility), and Fog (driving simulation with reduced visibility).

## Measurement Setup

The study is conducted in a simulated driving environment (Fig. 1) using a motion-based driving simulator at the Faculty of Electrical Engineering, University of Ljubljana, equipped with authentic vehicle components (seat, steering wheel, and pedals) and a physical dashboard. Simulation visuals are shown on three curved screens with full High Definition (HD) resolution (1920 x 1080 pixels), providing a 145° field of view of the driving environment. The driving scenario is created in SCANeR Studio (Stojmenova Pečečnik *et al.,* 2023).

The protocol is designed and conducted in accordance with the Declaration of Helsinki and Code of Ethics for researchers and the Guidelines for Ethical Conduct in Research Involving People issued by the University of Ljubljana (approval No. #049-2025). Prior to participation, written informed consent is obtained from all participants.

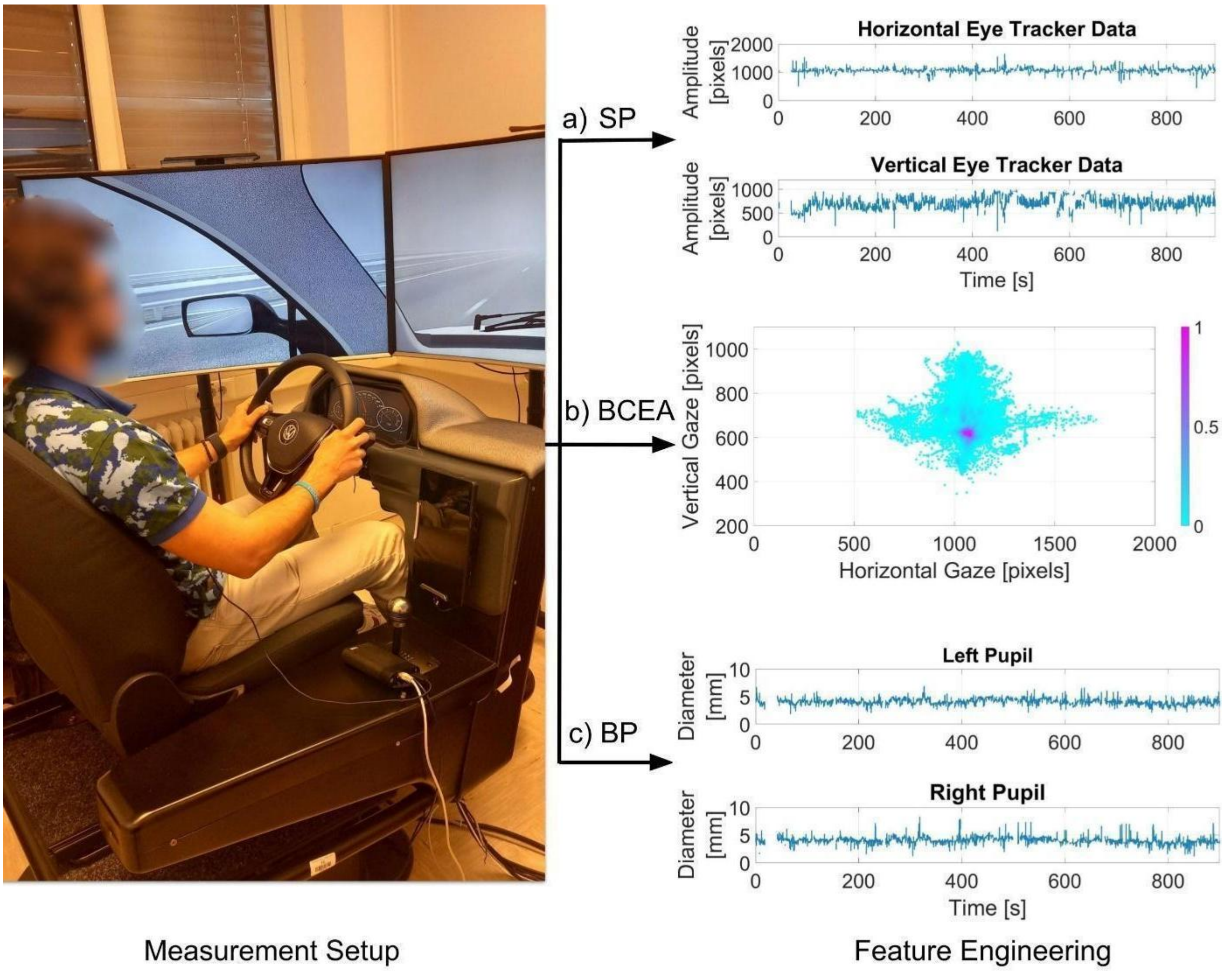

Fig. 1. Environment of a driving simulator and measured sample signals. The figure displays a glimpse of three groups of derived parameters from the eye tracking data: a) SP parameters derived from the presented raw signals of the horizontal and vertical degree of visual angle. b) BCEA parameters calculated from the scatter heat map of horizontal and vertical gaze point plots during the experiment; and c) BP parameters extracted from the diameter of the left and right pupils used for blink detection. SP - Saccade Parameters. BCEA - Bivariate Contour Ellipse Area. BP - Blinking Parameters.

After completing a demographic questionnaire, subjects are instructed to adjust the seat, as well as to mount and calibrate the eye tracker following the researchers' instructions. The entire measurement consists of three consecutive stages (Baseline, Ride, and Fog). During Baseline, participants are asked to sit still and relax for 15 minutes. After Baseline, participants familiarize themselves with the simulator setup, then drive on the highway (Ride). After ~10 minutes, weather conditions suddenly change, and dense fog appears along the highway, which lasts for around 2 minutes (Fig. 2). Participants' main task is only to drive safely and successfully reach the final destination.

## Data Acquisition

Overall, 28 participants (17 males and 11 females) aged from 21 to 42 (average age = 25.64, Standard Deviation (SD) = 5.76) with normal or corrected-to-normal vision, who hold a drivers' license (average year = 6.9, SD = 5.96), complete a task in a driving simulator. Participants' eye movements are recorded with the Tobii Pro Glasses 3 eye tracker (Tobii, 2025), with a sampling frequency of 100 Hz and stored for further analysis.

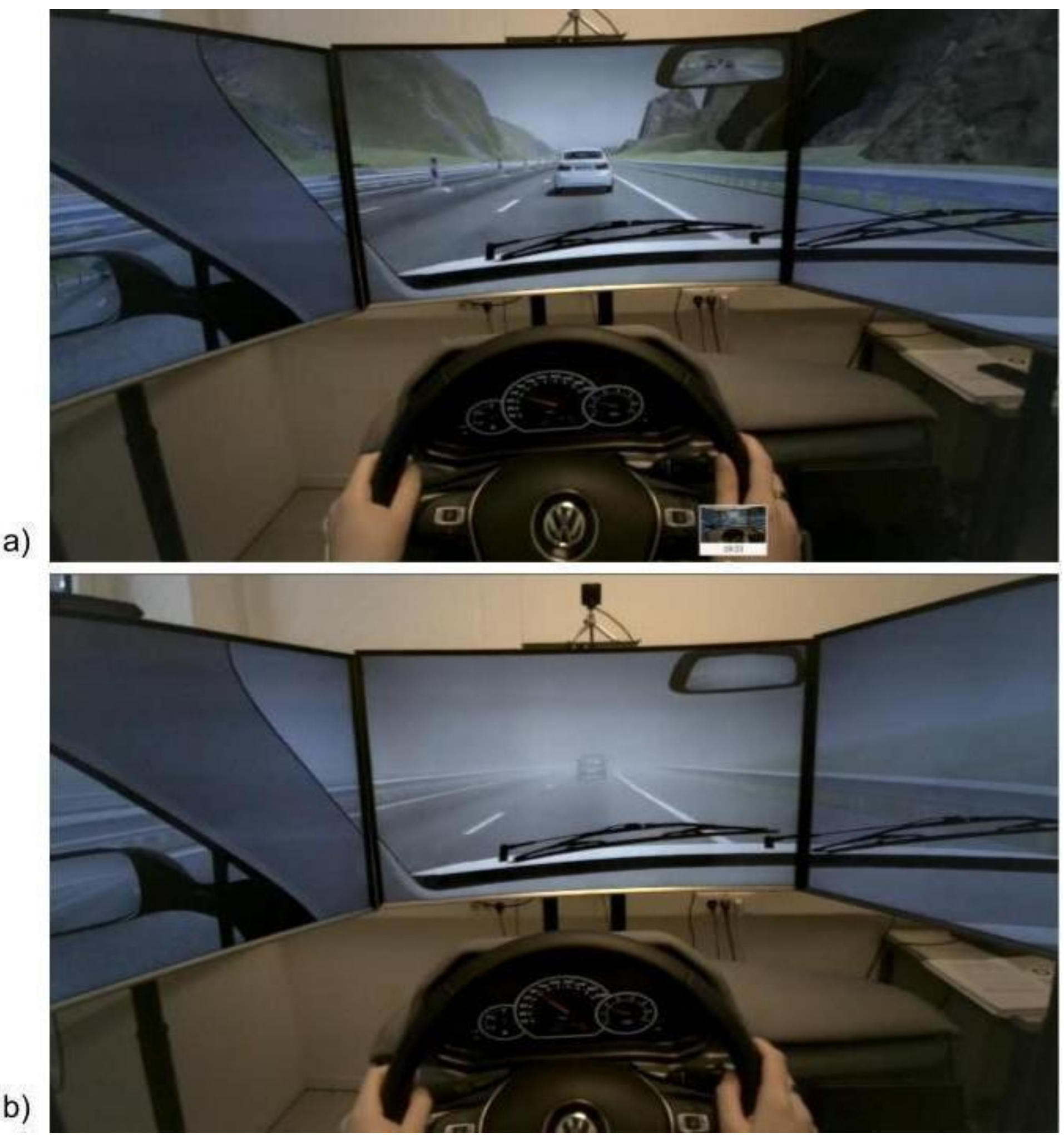

Fig. 2. Sample snapshots from the simulator. a) Driving during normal visibility (Ride). b) Driving during lower visibility (Fog).

## Preprocessing

All processing steps are performed in MATLAB R2021a (The Mathworks, Natick, USA), and the code for eye tracking analysis and feature extraction is freely shared with the anonymized dataset on GitHub and Zenodo (Stokanović *et al.,* 2026a; Stokanović *et al.,* 2026b). Due to the data loss (one participant has complete data loss, another has only baseline data, and two have partial recordings), the data of 4 participants are excluded, and the recordings obtained from 24 participants are analyzed. Normalized gaze coordinates that indicate the participants' gaze position as a proportion of the screen width and height are used and converted into pixel-based resolution (1920 x 1080).

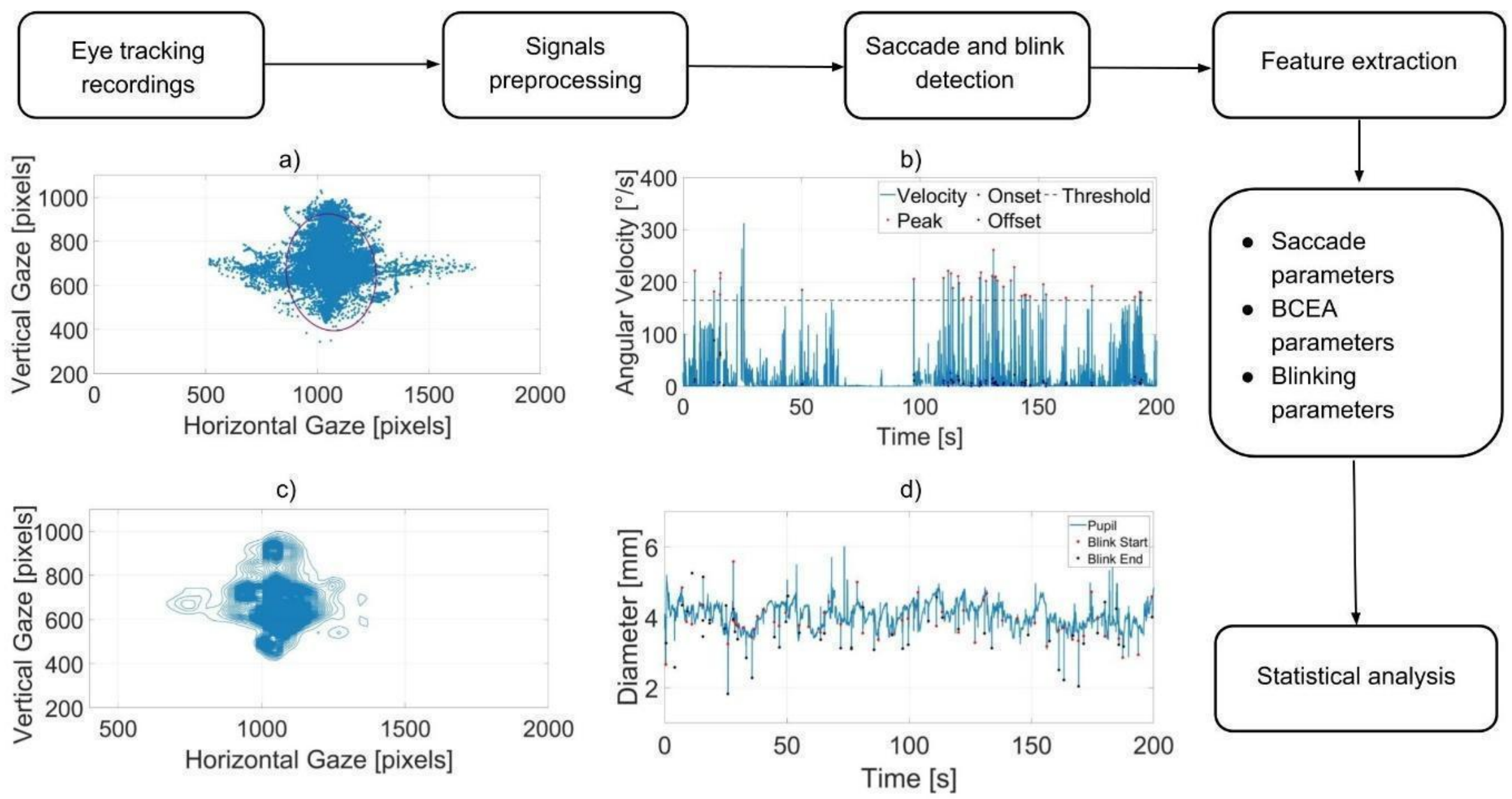

Fig. 3. Proposed methodological approach. a) Scatter plot of horizontal and vertical gaze point plots during the experiment with BCEA. b) Detected saccades on angular velocity profiles during the first 200 s of Baseline. c) Contour plots for detecting the number of Preferred Retina Locus (PRL). d) Detected blinks on the left pupil diameter during the first 200 s of Baseline. BCEA - Bivariate Contour Ellipse Area.

To reduce inaccuracies and distortions inherent in the eye tracking data (Niehorster *et al*., 2020a; Hessels *et al.,* 2016), missing or invalid samples are removed from the dataset to ensure continuity (Landes *et al.,* 2025). The remaining data are then smoothed using the third-order Savitzky-Golay filter (Savitzky and Golay, 1964), with an 11-sample window (110 ms at 100 Hz) (Zandi and Luhan, 2024) to reduce high-frequency noise while preserving the main signal features effectively, following approaches used in similar eye tracking studies (Schroeger *et al*., 2025). After preprocessing, the data are analyzed using the Identification - Velocity Threshold (I-VT) algorithm to extract relevant eye movement parameters, while blinks are identified based on signal loss in the pupil data, and BCEA related parameters are calculated (Fig. 3).

## Saccade Detection

Saccade detection is performed on angular velocity, calculated from the Cartesian gaze coordinates according to Eq. (1):

$$\dot{\theta} = \sqrt{(\phi_1 \dot{x})^2 + (\phi_2 \dot{y})^2} \quad (1)$$

where $\dot{x}$ and $\dot{y}$ represent Cartesian velocities (the first derivative of $x$ and $y$), while $\phi_1$ and $\phi_2$ represent pixel-to-angle transformation factors (95°/1920 and 63°/1080, respectively) (Tobii, 2025). Velocities higher than 1000°/s are considered to be physiologically impossible, so they are omitted from the analysis, as well as values lower than the signal median (Nyström and Holmqvist, 2010).

The adaptive I-VT algorithm improves upon the original I-VT algorithm by determining the optimal value of the multiplier parameter $n$ (Tanasković *et al*., 2024). The calculation of the adaptive threshold is shown in :

$$AT_{i+1} = mean(vel < AT_i) + n * sd(vel < AT_i) \quad (2)$$

where $AT_{i+1}$ represents the adaptive threshold in $i^{th}$ iteration, $sd$ standard deviation, $n$ represents the multiplier parameter, and $vel$ is the eye movement velocity profiles obtained by the central difference method (Nyström and Holmqvist, 2010; Tanasković *et al*., 2024). Optimization of the parameter $n$ begins by defining a range between 2.5 and 6, using a step of 0.5 (Tanasković *et al*., 2024). Afterwards, the optimal parameter $n$ is determined by calculating the misdetections of the saccades. As a criterion for calculating the misdetections, the psychological duration of the saccades is used as proposed in (Salvucci and Goldberg, 2000). Saccades with a duration shorter than 10 ms or longer than 100 ms are classified as misdetections (Nyström and Holmqvist, 2010; Tanasković *et al*., 2024; Imaoka *et al.,* 2020). In the Nyström-Holmqvist adaptive algorithm, the saccade peak velocity, $\theta_{PT}$ is set to an initial value within the range 100-300°/s, which we employ in this study. After setting to an initial value, a new threshold is calculated for all velocity samples lower than $\theta_{PT}$ as:

$$\theta_{PT} = \mu + \lambda\sigma \qquad (3)$$

where $\mu$ is the mean, $\sigma$ is the velocity standard deviation, and $\lambda$ is a scale factor set to 6. This procedure is then repeated until the error between iterations is less than 1°/s. To determine saccade onsets, the algorithm looks back in time from the putative saccade until it reaches the first point that crosses the saccade onset threshold defined in Eq. (4):

$$\theta_{ST}^{onset} = \mu + 3\sigma \qquad (4)$$

If the algorithm detects that a certain threshold has been crossed, it will go back in time to find the nearest local velocity minimum, which is considered the saccade onset. The process is similar for saccade offsets, except that the threshold for saccade offsets is defined differently:

$$\theta_{ST}^{offset} = 0.7\theta_{ST}^{onset} + 0.3LN \tag{5}$$

The velocity signal average and three times the standard deviation are calculated in the 40 ms before the saccade starts, and *LN* represents local noise, as presented by Nyström and Holmqvist (Nyström and Holmqvist, 2010), with the initial value for the threshold set to 200°/s. The initial value is set to 200°/s using error and trials. After saccade detection, numerous parameters are calculated as described in Table 1, including saccade amplitude, saccade duration, peak velocity, and fixation duration. For each subject, the mean, SD, and median of each parameter across all saccades are computed.

Table 1. Description of calculated saccade parameters

| Parameter | Explanation | Expectation | Reference(s) |
|---|---|---|---|
| Saccade amplitude [°] | The angle of eye movement from the beginning to the end of a saccade. | More complex driving situations, like navigating bends or tunnels, may lead to smaller saccade amplitudes as drivers focus on specific details. | (Guidetti *et al.*, 2019, Peng *et al.*, 2018; Han *et al.*, 2025) |
| Peak saccade velocity [°/s] | The maximum speed achieved during a saccade. | Peak velocities should be lower during mental attentional demand. | (Bachurina and Arsalidou, 2022; Di Stasi *et al.,* 2010*)* |
| Saccade duration [ms] | The duration of a saccade, measured from its initiation to its termination, typically ranges from 30 to 80 milliseconds. | Duration may be prolonged during driving and shorter than 200 ms. | (Guidetti *et al.* 2019*)* |
| Number of detected saccades [a. u.] | Quantification of detected saccadic eye movements. | An increased number of saccades is expected during driving and cognitive load. | (Biswas and Prabhakar, 2018) |
| Fixation duration [ms] | Duration spent on a single fixation point, approximately 200-300 ms. | Fixations during Baseline are expected to be longer and more stable, while driving may elicit shorter fixations due to the need for rapid visual sampling. | (Tullis and Albert, 2013; Recarte and Nunes, 2000; Foy and Chapman, 2018) |

## Computation of Bivariate Contour Ellipse Area Related Parameters

BCEA is proposed in Crossland *et al*. (2004) as a means to assess PRL quantitatively, calculated using Eq. (6):

$$BCEA = 2\pi k \sigma_x \sigma_y \sqrt{1 - \rho^2} \quad (6)$$

where $\sigma_x$ and $\sigma_y$ are the standard deviations of the horizontal and vertical eye movements, and $\rho$ is the Pearson product-moment correlation coefficient of the horizontal and vertical eye movements, with $k = 3$ corresponding to a probability value of 0.95 (Crossland *et al*., 2004). To assess fixation stability, we calculate the BCEA (Crossland *et al*., 2004; Upadhyaya *et al.,* 2017). In the context of driving, BCEA is valuable as it provides insights into visual attention and distraction. The present study represents the first application of BCEA metrics to characterize driver visual attention stability in dynamic driving scenarios. A smaller BCEA indicates more stable fixation and higher attention, while a higher BCEA represents increased wandering and potential distraction (Altemir *et al.,* 2024; Crossland *et al*., 2004; Moiroud *et al.,* 2025). The additional parameters shown in Table 2, including the mean and standard deviation of gaze points, correlation coefficient ($\rho$), and the number of PRL. GI, Mean Squared Error (MSE), and sample and approximate entropy, together, offer a comprehensive representation of fixation behavior and gaze stability.

Table 2. List of parameters for BCEA analysis

| Parameter | Explanation | Expectation | Reference(s) |
|---|---|---|---|
| Mean value of horizontal and vertical gaze data [°] | Average value of a participants' gaze along the horizontal and vertical axes. | During Baseline, horizontal and vertical positions are expected to remain close to the display center, while during driving, they are expected to increase. | (Foy and Chapman, 2018) |
| Standard deviation of horizontal and vertical gaze data [°] | Amount of variation in the horizontal and vertical position, indicating how much individual gaze positions deviate from the average position. | A higher standard deviation is expected during driving, especially for the horizontal gaze position. | Proposed here |
| $\rho$ [a.u.] | Pearson product-moment correlation coefficient of the horizontal and vertical eye movements. | $\rho$ should potentially increase during driving, due to coordinated diagonal gaze movements when scanning the roadway. | Proposed here |

| Parameter | Explanation | Expectation | Reference(s) |
|---|---|---|---|
| BCEA [$minarc^2$ ] | Bivariate Contour Ellipse Area | Higher BCEA is expected during Baseline, and lower BCEA during driving. | Proposed here anc previously used in clinical research (Crossland *et al*., 2004) |
| PRL [a.u.] | Preferred Retinal Locus | The number of PRL is expected to be higher while driving. | (Crossland *et al*., 2004) |
| Guzik's index (GI) [deg²] | Measurement of the asymmetry of BCEA | GI is expected to be higher during driving, reflecting more time spent above or below the major axis of the fixation ellipse. | Proposed here anc previously used fo Poincaré plot (Piskorski and Guzik, 2007; Karmakar *et al*., 2009; Karmakar *e* *al*., 2015) |
| MSE [deg²] | Mean Squared Error is used as a metric of the asymmetry of BCEA. | MSE is expected to be higher during driving, while fixations spread more around the ellipse major axis. | Proposed here |
| Sample and approximate entropy of horizontal and vertical gaze data [a. u.] | Measure of irregularity of gaze positions in horizontal and vertical data. | Higher values indicate greater irregularity, while lower values indicate more regularity. During driving, both measures are expected to decrease as visual attention becomes more focused. | Proposed here (Martinez-Cagig *e* *al*., 2018) |

Originally introduced in Electrocardiography (ECG) for analysis of heart rate variability, Guzik's index (GI) quantifies asymmetry in heart rate variability by comparing magnitudes of acceleration and deceleration in successive RR intervals (Karmakar *et al.,* 2009). In this study, GI is calculated to quantify the anisotropy of gaze dispersion relative to the major axes of the BCEA ellipse. First, gaze coordinates are centered by subtracting the mean horizontal and vertical positions from each sample. These centered data points are then projected on the major and minor axes of the BCEA ellipse to obtain the components of gaze variability parallel and perpendicular to the main direction of dispersion. GI is calculated as the ratio of the standard deviation along the major and along the minor axis, presented in Eq. (7):

$$GI = \frac{SD_{parallel}}{SD_{perpendicular}} \tag{7}$$

## Blink Detection

Besides saccades, blinking, and precise frequency of blinking represent crucial parameters to determine the level of fatigue and attention (Guidetti *et al.* 2019; Schleicher *et al.,* 2008*)*. Blink duration is expected to be between 150 ms and 500 ms, with 50 ms of fully closed eyes (Hollander and Huette*,* 2022; Hoppe *et al.*, 2018). For adults, the blink rate is between 16 and 20 blinks per minute (Hoppe *et al.*, 2018). Blink detection is performed on pupil diameter signals recorded at a sampling frequency of 100 Hz, following the procedure described by Nyström *et al.* (2024). To reduce the impact of prolonged signal loss due to artifacts unrelated to blinks, missing intervals longer than 200 consecutive samples (equivalent to 2 s at a 100 Hz sampling rate) are excluded from the analysis (Oliveira *et al.,* 2009; Grootjen *et al.*, 2023), while shorter gaps are preserved as potential blink candidates. A minimum blink duration of 100 ms and a maximum duration of 400 ms are applied to ensure physiological plausibility (Schiffman *et al.,* 1996; Demiral *et al.,* 2022). For each blink, the onset, the offset, and the blink duration are computed, as well as the blink rate. Parameters calculated for blink detection are described in Table 3.

## Statistical Analysis

A *post-hoc* power analysis is conducted using *G**Power (Faul *et al.,* 2009) for the repeated measures of Analysis of Variance (ANOVA) with three conditions (Baseline, Ride, and Fog). Using the observed average effect size across variables (Cohen's $d$ = 0.33) and within-subject correlation $\rho$ = 0.54, *G**Power estimated the statistical power to be 0.95, with an error probability of $\alpha$ = 0.05.

Table 3. Description of calculated parameters for blink detection

| Parameter | Explanation | Expectation | Reference(s) |
|---|---|---|---|
| Blink duration [ms] | The length of time during which the eyelids are closed during a blink | Longer blink durations are an indicator that a driver might be experiencing fatigue or increased workload. | (Goldstein *et al.,* 2007; Benedetto *et al.,* 2011; Ingre *et al.,* 2006) |
| Blink rate [a.u.] | Count of blinks detected during a specific period | Drowsiness can lead to a decreased blink rate during driving. | (Dari *et al.,* 2020) |
| Number of detected blinks per minute [a.u.] | Frequency of blinking in one minute | The average adult blinks around 15 - 20 times during one minute. | (Cornelis *et al.,* 2025) |

To assess differences in SP, BCEA-based parameters, and BP across conditions, Baseline, Ride, and Fog, statistical analysis is performed. All steps of statistical analysis are conducted using the Python 3.12 programming language (Rossum, 1995) and the Python code is shared on the GitHub repository (Stokanović *et al.,* 2026a). The following Python libraries are used for the realization of this research: NumPy (Harris *et al.,* 2020), Pandas (Pandas, 2025; McKinney and P.D.Team, 2025; McKinney, 2010), SciPy (Virtanen *et al.,* 2020), StatsModels (Seabold and Perktold, 2010), Seaborn (Waskom, 2021), Matplotlib (Hunter, 2007), Pingouin (Vallat, 2018), Scikit-Posthocs (Terpilowski, 2019), and IterTools (Python Itertools, 2025). During preprocessing, missing values are imputed using the median value (Han *et al.,* 2012). For each eye tracking metric, data across Baseline, Ride, and Fog conditions are tested for normality using the Shapiro-Wilk test. Parametric variables are analyzed using repeated measures ANOVA, while non-parametric variables are analyzed using the Friedman test. Obtained *p*-values are corrected using Benjamini-Hochberg false discovery rate (FDR) correction. Significant effects are followed by pairwise comparisons using paired t-tests (Cohen’s *d*) or Wilcoxon signed-rank tests (Cliff’s *δ*).

# Results

To get visual insight into the differences of Baseline, Ride, and Fog, boxplots are provided (Figs. 4-6) for all three groups of features: SP, BCEA-based parameters, and BP. In Figs. 4-6, separate colors are used to distinguish among three states: black for Baseline, blue for Ride, and purple for driving during Fog.

## Statistical Analysis of Saccade Parameters

The results of the statistical analysis for SP are presented in Table 4, which includes only statistically significant *post-hoc* comparisons ($p < 0.05$). The Friedman test reveals significant effects of condition for most parameters, indicating that saccade behaviour differs across Baseline, Ride, and Fog conditions. *Post-hoc* pairwise comparisons show that the mean and median saccade amplitudes are significantly greater during Ride and Fog compared to Baseline ($p < 0.01$; Cliff's $\delta$ = -0.19 to -0.34). Mean saccade duration is significantly greater in Baseline relative to Ride ($p < 0.01$; Cliff's $\delta = 0.38$).

Peak saccade velocity shows small differences between Ride and Fog ($p < 0.05$), suggesting that velocity dynamics remain relatively stable. However, fixation-related parameters demonstrate robust condition effects: mean and SD of fixation duration are significantly higher during Baseline compared to Ride and Fog ($p < 0.001$) with a very large effect (Cliff's $\delta = 0.74$). Median fixation duration follows a similar trend ($p < 0.01$), with a large effect (Cliff's δ = 0.62 - 0.64). The total number of saccades is reduced in Ride and Fog relative to Baseline ($p < 0.001$), indicating a very large effect size (Cliff's $\delta$ > 0.76).

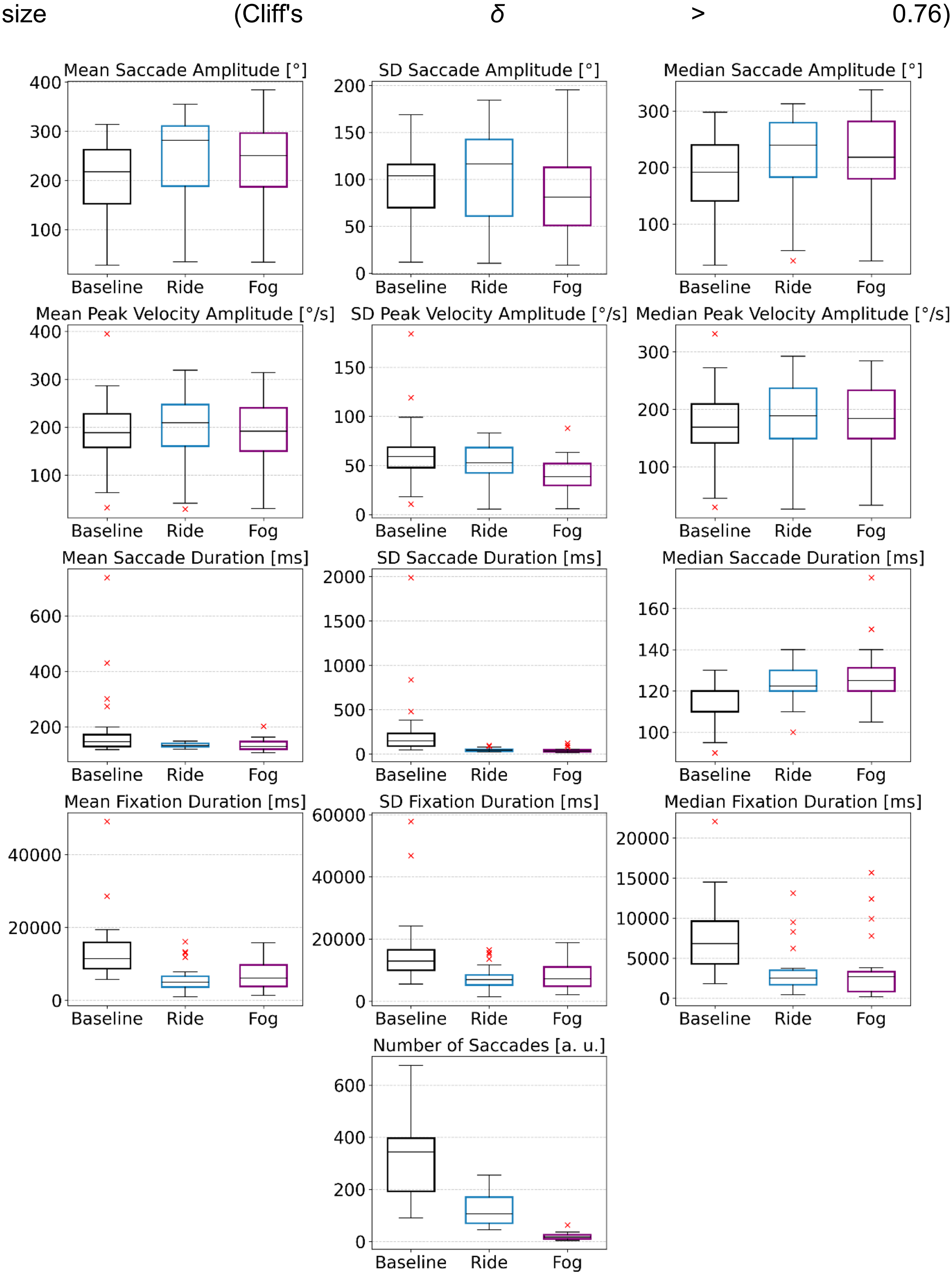

Fig. 4: Boxplots for saccade parameters during three conditions (Baseline, Ride, and Fog) recorded in the driving simulator. SD – Standard Deviation.

Table 4. Results of the statistical test for saccade parameters, significant difference *post-hoc* test (*p*-value > 0.05). SD – Standard Deviation. Statistically significant changes between Ride and Fog are marked in bold.

| Parameter | Condition | *p* – value | Effect | Effect Size | Interpretation |
|---|---|---|---|---|---|
| Mean Saccade Amplitude [°] | Baseline *vs* Ride | <0.001 | Cliff's *δ* | -0.34 | medium |
| | Baseline *vs* Fog | <0.01 | | -0.19 | small |
| SD Saccade Amplitude [°] | **Ride *vs* Fog** | **<0.05** | **Cohen's *d*** | **0.42** | **medium** |
| Median Saccade Amplitude [°] | Baseline *vs* Ride | <0.001 | Cliff's *δ* | -0.29 | small |
| | Baseline *vs* Fog | <0.001 | | -0.24 | small |
| Mean Peak Velocity Amplitude [°/s] | **Ride *vs* Fog** | **<0.05** | **Cohen's *d*** | **0.11** | **small** |
| SD Peak Velocity Amplitude [°/s] | Baseline *vs* Fog | <0.001 | Cliff's *δ* | 0.50 | large |
| | **Ride *vs* Fog** | **<0.01** | | **0.35** | **medium** |
| Mean Saccade Duration [ms] | Baseline *vs* Ride | <0.01 | Cliff's *δ* | 0.38 | medium |
| SD Saccade Duration [ms] | Baseline *vs* Ride | <0.001 | Cliff's *δ* | 0.87 | very large |
| | Baseline *vs* Fog | <0.001 | | 0.89 | very large |
| Median Saccade Duration [ms] | Baseline *vs* Ride | <0.01 | Cliff's *δ* | -0.57 | large |
| | Baseline *vs* Fog | <0.01 | | -0.55 | large |
| Mean Fixation Duration [ms] | Baseline *vs* Ride | <0.001 | Cliff's *δ* | 0.74 | very large |
| | Baseline *vs* Fog | <0.001 | | 0.63 | large |
| SD Fixation Duration [ms] | Baseline *vs* Ride | <0.001 | Cliff's *δ* | 0.66 | large |
| | Baseline *vs* Fog | <0.001 | | 0.57 | large |
| Median Fixation Duration [ms] | Baseline *vs* Ride | <0.001 | Cliff's *δ* | 0.64 | large |
| | Baseline *vs* Fog | <0.001 | | 0.62 | large |
| Number of Saccades [a. u.] | Baseline *vs* Ride | <0.001 | | 0.76 | very large |
| | Baseline *vs* Fog | <0.001 | Cliff's *δ* | 1.00 | very large |

| Parameter | Condition | *p* – value | Effect | Effect Size | Interpretation |
|---|---|---|---|---|---|
| | **Ride *vs* Fog** | **<0.001** | | **0.98** | **very large** |

## Statistical Analysis of Bivariate Contour Ellipse Area Features

Statistical results for BCEA related parameters are presented in Table 5 with statistically significant *post-hoc* comparisons ($p < 0.05$). The SD of horizontal decreases from Baseline to Ride($p < 0.001$; Cliff's $\delta = 0.24$) and shows a large effect for Baseline *vs* Fog and Ride *vs* Fog ($p < 0.001$; Cliff's $\delta = 0.53\text{-}0.59$), indicating that Fog decreases horizontal gaze variability. SD of vertical gaze, both Baseline *vs* Ride and Baseline *vs* Fog ($p < 0.001$; Cohen's $d = 1.34$ and $1.70$), with a medium effect between Ride and Fog, suggesting that vertical gaze variability changes moderately between the two driving conditions.

The correlation coefficient, $\rho$, shows a huge effect for both Baseline *vs* Ride and Baseline *vs* Fog ($p < 0.001$; Cohen's $d > 2.55$), indicating a strong difference in gaze correlation between these conditions. BCEA is significantly larger during Baseline than during driving conditions ($p < 0.001$), stipulating a substantial expansion of gaze dispersion during the Baseline. Significant differences are also found for the number of PRL in all comparisons (large effect), indicating a reduced number of PRL under Baseline and Ride conditions. GI is significantly higher in Baseline than in Ride and Fog ($p < 0.01$; Cohen's $d > 0.76$). MSE is significantly higher in Baseline than in Ride and higher in Baseline than in Fog ($p < 0.001$; Cliff's $\delta = 0.47\text{-}0.74$; very large effect), and higher in Ride than in Fog ($p < 0.001$; Cliff's $\delta = 0.51$; large effect), suggesting that gaze variability is lowest under the Fog simulation conditions. Vertical gaze entropy is significantly lower during Baseline and Ride than Fog ($p < 0.05$; Cliff's $\delta$ = -0.23 to -0.24), reflecting an increase in vertical gaze complexity under Fog.

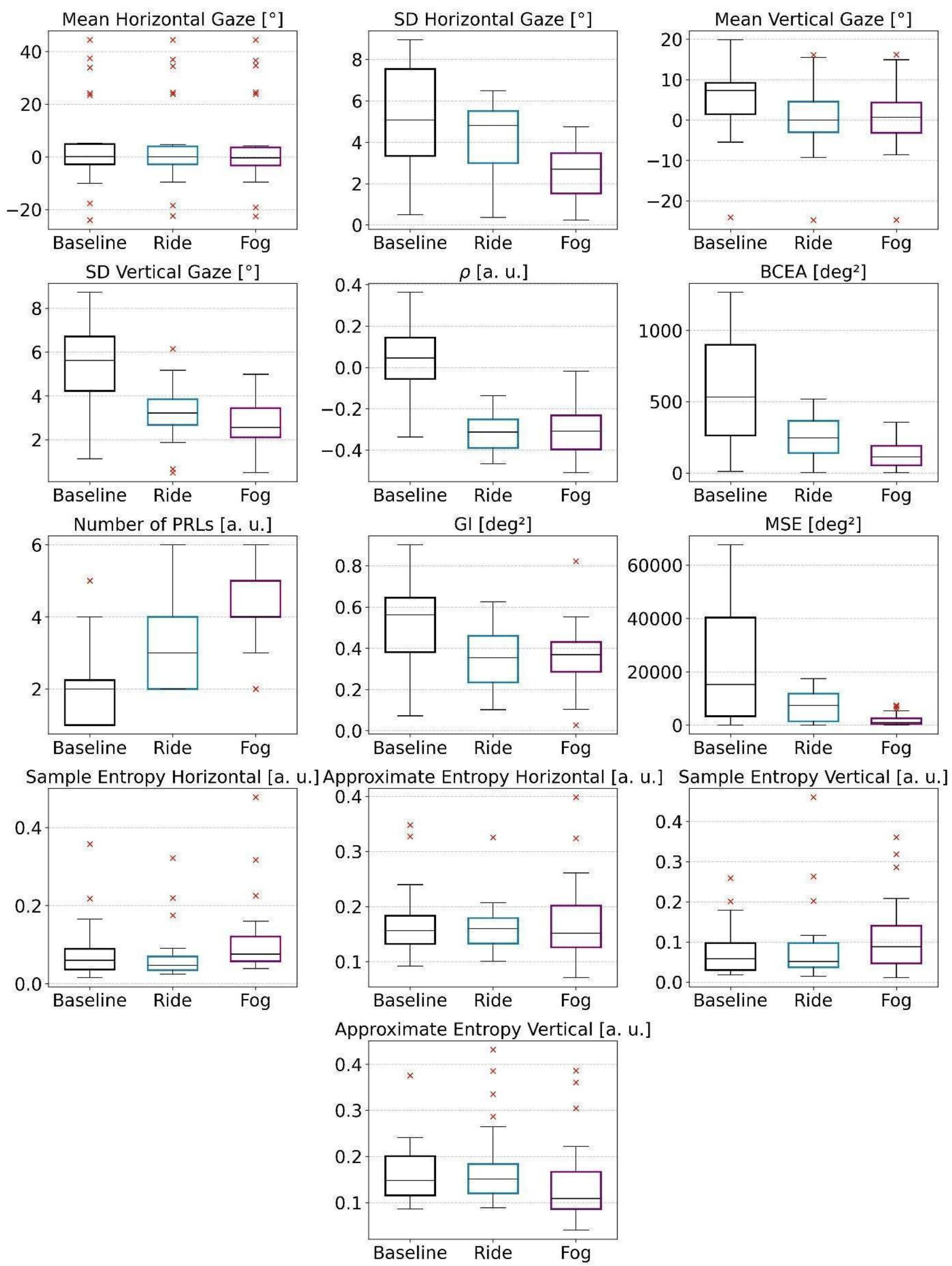

Fig. 5: Boxplots for BCEA parameters during three conditions (Baseline, Ride, and Fog) recorded within a driving simulator. SD – Standard Deviation, BCEA – Bivariate Contour Ellipse Area, PRL – Preferential Retinal Locus, GI - Guzik's Index, MSE - Mean Squared Error.

Table 5. Results of the statistical test for BCEA, significant difference *post-hoc* test (*p*-value > 0.05). SD – Standard Deviation, BCEA – Bivariate Contour Ellipse Area, PRL – Preferential Retinal Locus, GI - Guzik's Index, MSE - Mean Squared Error. Statistically significant changes between Ride and Fog are marked in bold.

| Parameter | Condition | *p* – value | Effect | Effect Size | Interpretation |
|---|---|---|---|---|---|
| Mean Horizontal Gaze [°] | **Ride *vs* Fog** | **<0.001** | **Cliff's δ** | **0.05** | **negligible** |
| SD Horizontal Gaze [°] | Baseline *vs* Ride | <0.001 | Cliff's δ | 0.24 | small |
| | Baseline *vs* Fog | <0.001 | | 0.59 | large |
| | **Ride *vs* Fog** | **<0.001** | | **0.53** | **large** |
| Mean Vertical Gaze [°] | Baseline *vs* Ride | <0.001 | Cliff's δ | 0.44 | medium |
| | Baseline *vs* Fog | <0.001 | | 0.44 | medium |
| SD Vertical Gaze [°] | Baseline *vs* Ride | <0.001 | Cohen's *d* | 1.34 | huge |
| | Baseline *vs* Fog | <0.001 | | 1.70 | huge |
| | **Ride *vs* Fog** | **<0.001** | | **0.42** | **medium** |
| $\rho$ [a. u.] | Baseline *vs* Ride | <0.001 | Cohen's *d* | 3.03 | huge |
| | Baseline *vs* Fog | <0.001 | | 2.55 | huge |
| BCEA [minarc²] | Baseline *vs* Ride | <0.001 | Cohen's *d* | 1.06 | very large |
| | Baseline *vs* Fog | <0.001 | | 1.53 | huge |
| | **Ride *vs* Fog** | **<0.001** | | **0.89** | **very large** |
| Number of PRL [a. u.] | Baseline *vs* Ride | <0.05 | Cliff's δ | -0.49 | large |
| | Baseline *vs* Fog | <0.01 | | -0.70 | large |
| | **Ride *vs* Fog** | **<0.05** | | **-0.49** | **large** |
| GI [deg²] | Baseline *vs* Ride | <0.01 | Cohen's *d* | 0.88 | very large |
| | Baseline *vs* Fog | <0.01 | | 0.76 | large |
| MSE [deg²] | Baseline *vs* Ride | <0.001 | | 0.47 | medium |

| Parameter | Condition | *p* – value | Effect | Effect Size | Interpretation |
|---|---|---|---|---|---|
| | Baseline *vs* Fog | <0.001 | Cliff's $\delta$ | 0.74 | very large |
| | **Ride *vs* Fog** | **<0.001** | | **0.51** | **large** |
| Sample Entropy Horizontal [a. u.] | Baseline *vs* Fog | <0.05 | Cliff's $\delta$ | -0.31 | small |
| | **Ride *vs* Fog** | **<0.001** | | **-0.46** | **medium** |
| Sample Entropy Vertical [a. u.] | Baseline *vs* Fog | <0.05 | Cliff's $\delta$ | -0.24 | small |
| | **Ride *vs* Fog** | **<0.05** | | **-0.22** | **small** |
| Approximate Entropy Vertical [a. u.] | **Ride *vs* Fog** | **<0.001** | Cliff's $\delta$ | **0.41** | **medium** |

## Statistical Analysis of Blink Parameters

Blinking behaviour is analyzed to assess changes in ocular activity during Baseline, Ride, and Fog conditions (Table 6). Blink duration decreases substantially during Ride and Fog compared to Baseline: mean ($p < 0.001$; Cliff's $\delta > 0.90$), median ($p < 0.05$; Cliff's $\delta > 0.70$), and SD of blink duration ($p < 0.01$; Cliff's $\delta > 0.80$) all show very large effects. Blink rate is significantly higher in Ride and Fog than Baseline ($p < 0.001$; Cliff's $\delta$ = 0.95-0.99; very large effect), and medium effect comparing Ride and Fog ($p < 0.001$, Cliff's $\delta = 0.39$). Number of blinks per minute is elevated in Ride and Fog *versus* Baseline ($p < 0.001$; Cliff's $\delta$ = 0.89-0.91; very large effect), with a small, but statistically significant difference between Ride and Fog ($p < 0.01$; Cliff's $\delta = 0.39$).

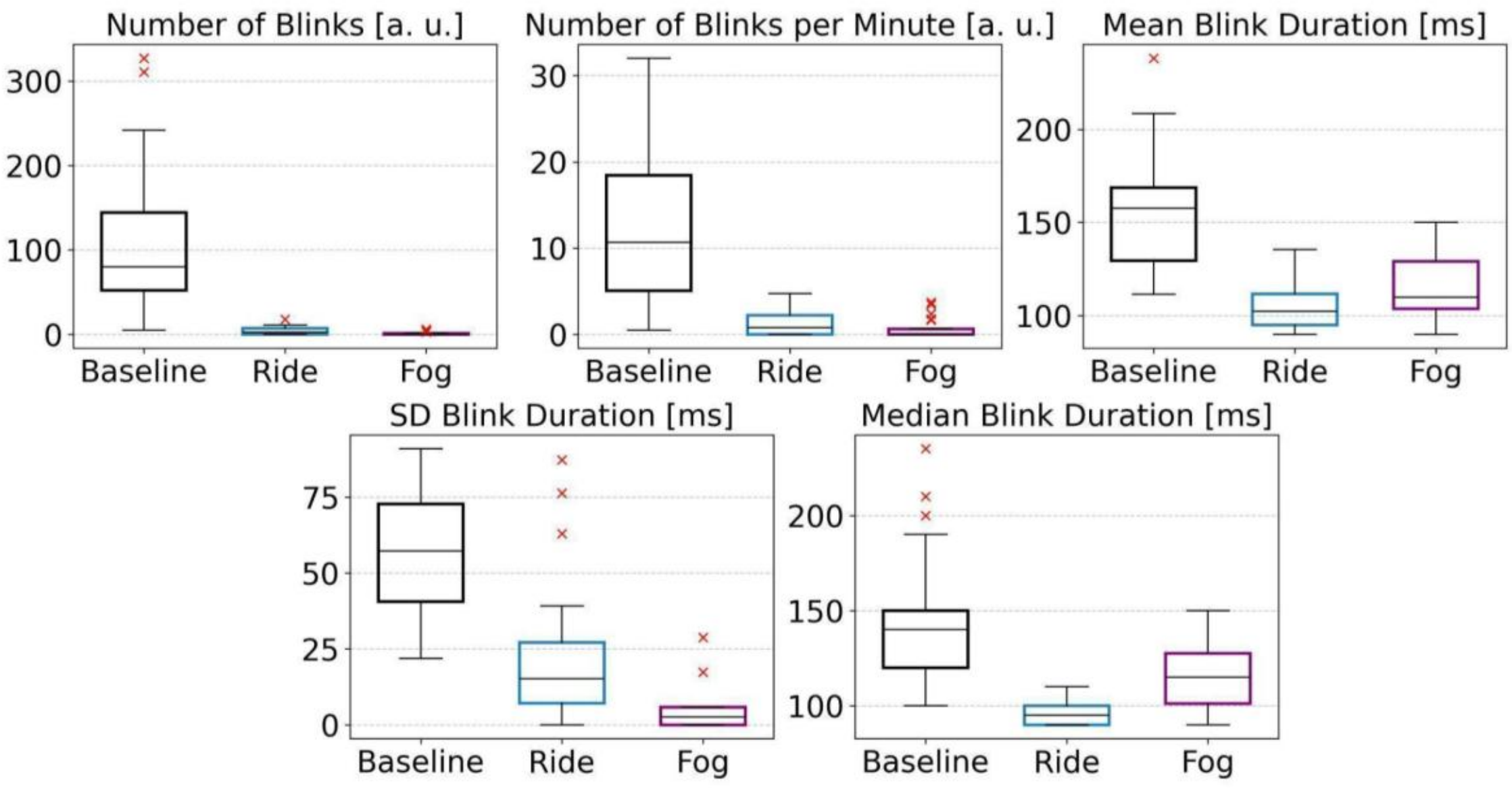

Fig. 6: Boxplots for blinking parameters during three conditions (Baseline, Ride, and Fog) recorded within a driving simulator. SD – Standard Deviation.

Table 6. Results of the statistical test for blinking parameters, significant difference *post-hoc* test ($p$-value > 0.05). SD – Standard Deviation. Statistically significant changes between Ride and Fog are marked in bold.

| Parameter | Condition | *p* – value | Effect | Effect Size | Interpretation |
|---|---|---|---|---|---|
| Mean Bllink Duration [ms] | Baseline *vs* Ride | <0.01 | Cliff's δ | 0.95 | very large |
| | Baseline *vs* Fog | <0.01 | | 0.90 | very large |
| SD Blink Duration [ms] | Baseline *vs* Fog | <0.01 | Cliff's δ | 1.00 | very large |
| | **Ride *vs* Fog** | **<0.01** | Cliff's δ | **0.80** | **very large** |
| Median Blink Duration [ms] | Baseline *vs* Ride | <0.01 | | 1.00 | very large |
| | Baseline *vs* Fog | <0.01 | Cliff's δ | 0.70 | large |
| | **Ride *vs* Fog** | **<0.05** | | **-0.68** | **large** |
| Blink Rate [a.u.] | Baseline *vs* Ride | <0.001 | | 0.95 | very large |
| | Baseline *vs* Fog | <0.001 | Cliff's δ | 0.99 | very large |
| | **Ride *vs* Fog** | **<0.001** | | **0.39** | **medium** |

| Parameter | Condition | *p* – value | Effect | Effect Size | Interpretation |
|---|---|---|---|---|---|
| Number of Detected Blinks per Minute [a.u.] | Baseline *vs* Ride | <0.001 | Cliff's *δ* | 0.88 | very large |
| | Baseline *vs* Fog | <0.001 | | 0.91 | very large |
| | **Ride *vs* Fog** | **<0.01** | | **0.32** | **small** |

# Discussion

The present study examines oculomotor behavior across different conditions by analyzing three groups of features obtained from eye movement measurements: SP, BCEA-based parameters, and BP.

## Insights into Saccadic Eye Movement Dynamics

Saccade amplitudes increase during Ride and Fog compared to Baseline, contrary to the expectation that a more demanding visual environment leads to smaller gaze shifts (Guidetti *et al.,* 2019; Peng *et al.,* 2018; Han *et al.,* 2025). These results suggest that under increased visual and cognitive load, participants perform larger gaze shifts, probably with the aim of extracting more information from the environment. Similar findings were reported during distracted or multitasking states (Wang *et al.,* 2023; Lee *et al.,* 2024) and in studies where more visually complex environments prompted more extensive saccades (Zhao *et al.,* 2023). Mean saccade duration decreases during the Ride, which does not align with the prediction that saccades would become slightly longer under higher visual and cognitive load (Lee *et al.,* 2024). Peak saccade velocity shows only small differences between conditions ($p < 0.05$; Cliff's $\delta = 0.11$), consistent with prior work suggesting velocity changes modestly with task difficulty in driving (Galley, 1993; Gorin *et al.*, 2024).

Fixation duration decreases substantially during both driving conditions (very large and large effect, Cliff's $\delta > 0.63$), reflecting shorter, more frequent fixations. This aligns with HCI research showing that task complexity and environmental demands modulate fixation behaviour to optimize information acquisition (Goldberg and Kotval, 1999; Chen *et al.,* 2011; Liu *et al.,* 2022). The total number of saccades is also markedly reduced during Ride and Fog (Cliff's $\delta$ = 0.76-1.0), being consistent with findings from tunnel driving studies (Wang *et al.,* 2016).

## Is Gaze Variability a Reliable Indicator of Visual Attention?

To further characterize gaze behaviour, calculation of BCEA-based parameters is presented during Baseline and driving conditions in Table 5. Parameters such as mean vertical gaze and SD of horizontal and vertical gaze decrease during driving conditions, indicating a larger, wider dispersion across the visual field during Baseline. Horizontal gaze variability shows a large effect for Baseline *vs* Fog and Ride *vs* Fog (Cliff's $\delta$ = 0.53-0.59), indicating that Fog strongly narrows horizontal scanning. Vertical gaze variability shows a large effect across all Baseline comparisons (Cohen's $d$ = 1.34-1.70), with a medium effect between Ride and Fog, which is in line with prior findings that vertical is more sensitive to environmental demands than horizontal gaze (Halin *et al.,* 2025; Louw and Merat, 2017). Previous studies showed that drivers fixate closer to the vehicle during night time driving compared to daytime conditions (Graf and Krebs, 1976; Rockwell, 1970). Our results under Fog conditions show a similar pattern, with vertical gaze becoming more focused when visual information is limited, suggesting that drivers adaptively narrow their gaze under challenging conditions.

The correlation coefficient, $\rho$, and BCEA indicate a significant effect between conditions, with a decrease during driving conditions, reflecting a more focused visual strategy when participants face higher task demands**.** BCEA decreases from Baseline to Ride (Cohen's $d$ = 1.06) and from Baseline to Fog (Cohen's $d$ = 1.53), confirming more stable, focused fixations under task engagement (Crossland *et al*. 2004; Kim *et al*., 2022). The significant difference between Ride and Fog (Cohen's $d$ = 0.89) indicates that gaze stability increases as visibility degrades, suggesting a progressive oculomotor adaptation to worsening visibility.

The number of PRL increases during driving conditions, while participants focus on the road, rearview mirrors, and dashboard ($p < 0.05$, Cliff's $\delta$ = -0.49 and -0.70; large effect) reflecting more distributed fixation strategies during active environmental monitoring, consistent with functional gaze strategies under challenging visual conditions, such as during central vision loss or increased task demands (Bronstad *et al.,* 2015; Kujala *et al.*, 2016). This confirms that during driving, participants are engaged in more active visual exploration across multiple areas of interest, such as the road ahead, rearview mirrors, and the dashboard.

Analysis of GI and MSE reveals significant changes in fixation asymmetry. Both parameters are higher during Baseline compared to Ride and Fog, indicating that fixations are more widely dispersed under resting or low-demand conditions. Specifically, GI is significantly higher during Baseline than both driving conditions ($p$ < 0.01, Cohen's $d$ = 0.76-0.88; very large). MSE decreases progressively across conditions and, importantly, distinguishes Ride from Fog (Cliff's $\delta$ = 0.51), suggesting that gaze dispersion is sensitive not only to Baseline *versus* driving, but also to differences between driving conditions.

Entropy analyses indicate that gaze patterns become less predictable under increased task demands. Vertical gaze complexity increases under Fog (Cliff's $\delta$ = - 0.23 to -0.24), while approximate entropy decreases from Ride to Fog (Cliff's $\delta$ = 0.41), reflecting that eye movements become less variable under limited visibility. These findings align with research demonstrating that cognitively demanding tasks produce more concentrated and predictable scanning patterns (Cui *et al.*, 2024; Krejtz *et al.*, 2014; Krejtz *et al.,* 2015). This suggests that entropy measures capture adaptive visual strategies in response to environmental challenges.

## What do Blinks Really Reveal?

Mean and median values of blink duration decrease during driving conditions compared to Baseline, as well as SD, while blink rate and number of blinks per minute increase. This pattern (shorter, more frequent, and more consistent blinks) indicates that environmental challenges, such as motion and reduced visibility, lead to shorter, more consistent, and less frequent blinks, suggesting a reduction in blinking activity, possibly to maintain visual stability under demanding conditions. Blink duration was a more reliable indicator of visual workload than blink rate in driving contexts (Benedetto *et al.,* 2011) and cognitive and combined task loads have been associated with an increased blink rate and altered gaze dynamics (Liang and Lee, 2010; Ayala *et* al., 2024).

Published findings support blink suppression pattern, showing that increased visual workload is associated with shorter blinks (71 - 100 ms; $p < 0.05$) while longer blinks (171-300 ms; $p < 0.05$) occur more frequently as time-on-task increases, reflecting fatigue-related changes in oculomotor control (Benedetto *et al.,* 2011). Bachurina and Arsalidou (2022) found a significant main effect of mental attentional demand on spontaneous blink rate, with reduced blinking for higher demand levels. Studies of visually demanding tasks - from video games Argilés *et al.,* (2025) to display terminal work (Cardona *et al.*, 2011), where blink rate dropped to nearly one-third of baseline (Cardona *et al.,* 2011), demonstrated strategic blink suppression to maximize visual information acquisition (Orchard and Stern, 1991). Our findings align with this pattern: reduced visibility in fog elicits similar oculomotor adaptations to those observed in other high-workload scenarios, further validating blink measures as sensitive indicators of driver cognitive state.

## The Drivers' Adaptive Gaze Toolkit

Overall, the present findings indicate that drivers adapt their oculomotor behavior to cope with increased visual and cognitive demands. Under Ride and Fog conditions, saccade amplitude increases, while saccade number decreases, reflecting more targeted gaze shifts to extract critical information from the environment. BCEA measures show reduced gaze dispersion, particularly vertically, suggesting a more focused and efficient visual strategy. Such adaptation likely reflects an effort to maintain stability and optimize visual processing under challenging sensory conditions. BP further supports this adaptive behaviour, with shorter, less frequent, and more consistent blinks under higher workload conditions, highlighting the sensitivity of blink measures to cognitive load. Altogether, drivers dynamically modulate saccadic and fixation behavior, gaze distribution, and blinking patterns to maintain visual stability when facing environmental complexity.

From an HCI perspective, the single sensor, multiparameter methodology provides an objective way to evaluate how interfaces perform when visibility is reduced. Our results suggest that interfaces should adapt to environmental conditions when detecting high workload patterns (reduced gaze dispersion or fewer blinks) in a way to increase focus: systems could simplify displays, increase text size, or delay secondary alerts. This supports the development of automotive interfaces that respond to both driver state and comfort, as well as to environmental conditions.

## Limitations and Future Research

This study is not without limitations:

1. Sampling frequency influences the accuracy of eye movement detection (Imaoka *et al.,* 2020; Leube *et al.,* 2017). Systems with higher frequencies (>250 Hz) are recommended to ensure precise measurement. Although our 100 Hz system operated below this threshold, established preprocessing, interpolation, or filtering approaches can further improve data quality (Tanasković *et al*., 2024; Antoniades *et al.,* 2013; Wierts *et al.,* 2008).

2. Using an eye tracker is also a notable limitation (Holmqvist *et al.,* 2012). Despite their widespread usage, recent systematic reviews of eye tracking have highlighted ongoing challenges in data quality, cross-study comparability, and the need for standardized protocols across different eye tracking hardware platforms (Dunn *et al*., 2023; Jakobi *et al*., 2024): data loss (Niehorster *et al*., 2020b), slippage (Bulling *et al.,* 2009), scene instability (Evans, 2004), gaze accuracy issues (Niehorster *et al*., 2020b), lighting variations (Bednarik *et al*., 2012).

3. Although driving simulators offer a safe and controlled driving experience (Evans, 2004; Stein and Dubinsky, 2011; Himmels *et al.,* 2024), there is always a lack of realistic physical sensation, weather (Chan *et al*., 2010), and potential for simulator sickness (de Winter *et al.,* 2012; Jakus *et al.,* 2022).

4. A direct comparison of the exact metric values across more diverse driving simulations should be taken with a grain of salt, as variations in driving scenarios, such as road geometry (straight highways *vs.* curved urban streets), environmental complexity (sparse rural *vs.* dense city traffic), and visibility conditions (clear *vs.* fog), systematically influence gaze behavior. Meaningful comparisons require standardized scenarios or focus on relative rather than absolute cross-study values.

5. Machine learning can be used for saccade detection and classification of eye tracking data into saccade or fixation (Jothi Prabha and Bhargavi, 2020; Fridman *et al.,* 2018; Zemblys *et al.,* 2018; Akshay *et al.,* 2020; Rizzo *et al.,* 2022; Liu *et al.,* 2021).

6. Cognitive and combined loads have been associated with greater blink rates and altered gaze patterns (Liang and Lee, 2010). Integrating such conditions into the simulator and HCI designs could provide deeper insight into driver attention and workload.

7. Future directions could explore analysis of the eye openness signal to improve blink and fatigue detection (Nyström *et al.,* 2024; Häkkänen *et al.,* 1999; Shekari Soleimanloo *et al.,* 2019).

8. Implementing additional algorithms for saccade detection, such as the Median Absolute Deviation (MAD) algorithm (Voloh *et al.,* 2020) or dispersion-based algorithms (Salvucci and Goldberg, 2000), may enhance the identification robustness. However, methodological challenges for blink detection, including the lack of standardized approaches for dealing with missing data, artifacts elimination, and the need for consistent data preprocessing pipelines across studies should be taken into account (Grootjen *et al.*, 2024).

9. Context-sensitive distraction warning systems, which adapt to upcoming traffic demands (Kujala *et al.*, 2024), could leverage the oculometric patterns identified in our study (increased entropy during mind-wandering and decreased blink rate during high-demand tasks) to provide just-in-time warnings that are neither premature or delayed.

# Conclusion

In this study, eye tracking data are used to investigate visual attention and gaze stability during Baseline and simulated driving conditions with varying visibility. Together with saccade and blinking parameters, the BCEA measures provide complementary insights into gaze dispersion and visual focus. The results demonstrate that reduced visibility induces pronounced changes in oculomotor behavior, including fewer but larger saccades, decreased gaze dispersion, and altered blinking patterns. The inclusion of the novel parameter, GI, further enhances the sensitivity of eye movement analysis, capturing fixation asymmetry not reflected in conventional metrics. Overall, these findings underscore the value of combining saccade, BCEA, and blink parameters for comprehensive assessment of visual attention and workload, offering potential applications for driving safety evaluation and automotive HCI design with a multiparameter and single sensor approach.

## Declaration of Competing Interest

The Authors declare that they have no known competing financial interests or personal relationships that could have appeared to influence the work reported in this paper.

## CRedit Authorship Contribution Statement

Smilja Stokanović: Writing - original draft, Visualization, Software, Methodology, Formal analysis, Data curation, Conceptualization. Jaka Sodnik: Writing

- review & editing, Methodology, Formal analysis, Data curation, Nadica Miljković: Writing - review & editing, Validation, Methodology, Investigation, Conceptualization.

## Acknowledgment

We express our deep gratitude to Jelena Medarević, from the Faculty of Electrical Engineering, University of Ljubljana, and Đorđe D. Nešković, from the School of Electrical Engineering, University of Belgrade, and Vinča Institute of Nuclear Sciences-National Institute of the Republic of Serbia, University of Belgrade.

N. M. kindly acknowledges the support from the Grant No. 451-03-34/2026-03/200103 funded by the Ministry of Science, Technological Development, and Innovation of the Republic of Serbia.

This work has also been financially supported by the European Union's Horizon Europe research and innovation program for the project FRODDO, grant agreement no. 101147819 and by the Slovenian Research and Innovation Agency within the program ICT4QL, grant No. P2-0246.

## Data and Code Availability

The source code for eye tracking analysis, including I-VT algorithm implementation, BCEA calculation, entropy measures, and blink parameter extraction, is available and shared under the GNU GPL-3.0 license on GitHub with assigned Zenodo permanent identifier (Stokanović *et al.*, 2026a). An anonymized dataset comprising gaze position is shared openly via Zenodo with Creative Commons Attribution 4.0 International license CC BY (Stokanović *et al*., 2026b). This enhances reusability and reproducibility, while enabling the community to extend our methodology to related attention-aware interactive systems.

## Statement

During the writing process, the Authors used GPT4o (ChatGPT) and Grok (model 4) to enhance readability and refine language. The Authors subsequently reviewed and revised the content as needed and take full responsibility for the content of the publication.